\documentclass[twocolumn,prd,amsmath,superscriptaddress]{revtex4}
\usepackage{graphicx}
\usepackage{latexsym}
\usepackage{amsmath}
\usepackage{amssymb}
\usepackage{graphics}
\usepackage{color}
\usepackage{hyperref}
\usepackage{bm}

\newcommand{\bbr}{\beta_\mathrm{br}}
\newcommand{\bgbr}{\beta_{\gamma\mathrm{,br}}}
\newcommand{\dVeff}{V_{\mathrm{eff},\phi}}
\newcommand{\ddVeff}{V_{\mathrm{eff},\phi\phi}}
\newcommand{\DES}{\Delta E_\mathrm{Sh}}
\newcommand{\DET}{\Delta E_\mathrm{Th}}
\newcommand{\Dzmin}{\Delta z_\mathrm{min}}
\newcommand{\DzST}{\Delta z_\mathrm{S-T}}
\newcommand{\eotwash}{E\"ot-Wash}
\newcommand{\ES}{E_\mathrm{Sh}}
\newcommand{\ET}{E_\mathrm{Th}}
\newcommand{\fscreen}{f_\mathrm{scr}}
\newcommand{\gcmc}{g/cm${^3}$}
\newcommand{\kg}{k_\mathrm{g}}
\newcommand{\mbr}{m_\mathrm{br}}

\newcommand{\Mg}{M_\gamma}

\newcommand{\Mpl}{M_\mathrm{Pl}} 
\newcommand{\muv}{\mu_\mathrm{v}} 
\newcommand{\Nholes}{N_\mathrm{h}}
\newcommand{\Nrows}{N_\mathrm{r}}
\newcommand{\phibr}{\phi_\mathrm{br}}
\newcommand{\phig}{\phi_\mathrm{g}}
\newcommand{\phis}{\phi_\mathrm{s}}
\newcommand{\rhom}{\rho_\mathrm{m}} 
\newcommand{\rhov}{\rho_\mathrm{v}}
\newcommand{\rS}{r_\mathrm{Sh}}
\newcommand{\rT}{r_\mathrm{Th}}
\newcommand{\varphis}{\varphi_\mathrm{s}}
\newcommand{\Veff}{V_\mathrm{eff}} 
\newcommand{\zfoil}{z_\mathrm{foil}}

\begin{document}

\title{Symmetron dark energy in laboratory experiments}
\author{Amol Upadhye}
\affiliation{High Energy Physics Division, Argonne National Laboratory, 9700 S. Cass Ave., Argonne, IL 60439}%

\date{\today}

\begin{abstract}
The symmetron scalar field is a matter-coupled dark energy candidate which effectively decouples from matter in high-density regions through a symmetry restoration.  We consider a previously unexplored regime, in which the vacuum mass $\mu \sim 2.4\times 10^{-3}$~eV of the symmetron is near the dark energy scale, and the matter coupling parameter $M \sim 1$~TeV is just beyond Standard Model energies.  Such a field will give rise to a fifth force at submillimeter distances which can be probed by short-range gravity experiments.  We show that a torsion pendulum experiment such as \eotwash~can exclude symmetrons in this regime for all self-couplings $\lambda \lesssim 7.5$.
\end{abstract}

\maketitle


\section{Introduction}
\label{sec:introduction}

Evidence for a large-scale acceleration of the cosmic expansion is now solid~\cite{Komatsu_etal_2010,Larson_etal_2011,Suzuki_etal_2012,Sanchez_etal_2012}, but we have yet to determine its cause.  Current data are consistent with a ``cosmological constant'' vacuum energy density as well alternative explanations known as ``dark energy.''  
Qualitatively, dark energy may differ from a cosmological constant in one of two ways.  Its energy density may evolve by a factor of order unity at recent times, or it may couple to known particles more strongly than gravity, provided that the resulting fifth forces are screened locally.

For scalar fields, the simplest dynamical dark energy models, several screening mechanisms have been found to evade laboratory and solar system searches for fifth forces.  In a chameleon model~\cite{Khoury_Weltman_2004a,Khoury_Weltman_2004b,Brax_etal_2004}, a nonlinear self-interaction makes the effective mass of the field large in high-density environments, reducing the length over which the fifth force operates.  A Galileon field~\cite{Vainshtein_1972,Dvali_Gabadadze_Porrati_2000,Deffayet_etal_2001,Nicolis_Rattazzi_Trincherini_2008} has a non-canonical kinetic energy, causing it to decouple from matter at high densities.  The symmetron is a canonical scalar whose effective potential is symmetric under $\phi \rightarrow -\phi$~\cite{Hinterbichler_Khoury_2010,Olive_Pospelov_2007}.  In regions of low density, the symmetry is spontaneously broken, and the effective couplings are proportional to the vacuum expectation value (VEV).  At high density the symmetry is restored, the VEV becomes zero, and couplings vanish.

If the density of a laboratory vacuum is low enough for the symmetron field to enter its broken-symmetry phase, then the field will mediate a fifth force between massive objects in that vacuum, which may be probed experimentally~\cite{Adelberger_Heckel_Nelson_2003,Adelberger_etal_2009,Upadhye_Gubser_Khoury_2006,Brax_etal_2007c,Brax_etal_2010b}.  For a symmetron mass $\mu \sim 10^{-3}$~eV  and a matter coupling energy $M \sim 1$~TeV, this symmetry breaking will occur at densities $\rho < \mu^2 M^2 \sim 0.1$~\gcmc~and distances $\sim \mu^{-1} \sim 0.1$~mm readily accessible to short-range gravity experiments such as the \eotwash~torsion pendulum~\cite{Kapner_etal_2007}.  
In this work we solve the symmetron equations of motion exactly in a planar geometry.  Using this result, we generalize the one-dimensional plane-parallel (1Dpp) approximation of~\cite{Upadhye_2012a} to symmetron models, allowing us to estimate torsion pendulum constraints. In particular, we show that all self-couplings $\lambda < 7.5$ are excluded for $M=1$~TeV and $\mu$ at the dark energy scale.  A very interesting region of parameter space is accessible to existing experiments. 

This article is organized as follows.  In Sec.~\ref{sec:symmetron_phenomenology} we describe the symmetron mechanism and apply it to one-dimensional matter configurations.  Section~\ref{sec:constraints_on_symmetrons} uses the 1Dpp approximation to estimate constraints on symmetrons, and Sec.~\ref{sec:conclusion} concludes.


\section{Symmetron phenomenology}
\label{sec:symmetron_phenomenology}

{{\bf{\em{A.~Symmetrons and screening: }}}
For nonrelativistic matter in a flat spacetime background, 
the symmetron action is approximated by
\begin{equation}
S
=
\int d^4x  \left[  
  -\frac{(\partial \phi)^2}{2}
  - \left( \frac{\rho}{M^2} 
  - \mu^2\right) \frac{\phi^2}{2}
  - \frac{\lambda}{4!}\phi^4
  \right]
\label{e:action}
\end{equation}
where $\rho$ is the matter density, $\lambda$ is the dimensionless self-coupling, and both the coupling energy $M$ and the vacuum mass $\mu$ have units of energy.  The vacuum energy resulting from~(\ref{e:action}) by itself cannot lead to the observed acceleration, so we must add a constant term $M_\Lambda^4$, where $M_\Lambda = 2.4\times 10^{-3}$~eV is the dark energy scale; however, such a constant is invisible in laboratory experiments.

Previous constraints have considered $\mu \sim 10^3 M_\Lambda^2/\Mpl$ and $M$ around the GUT scale, leading to unscreened fifth forces on cosmological scales~\cite{Hinterbichler_Khoury_2010,Hinterbichler_Khoury_Levy_Matas_2011}.  Here we are interested in $\mu \sim M_\Lambda$ and $M \sim 1$~TeV, an energy possibly associated with physics beyond the Standard Model.  We will see that such scales result in new effects in laboratory experiments.
The symmetron equation of motion is
\begin{equation}
\Box \phi
=
\dVeff(\phi,\rho),
\quad
\Veff
=
\left( 
\frac{\rho}{M^2} 
- \mu^2
\right) \frac{\phi^2}{2}
+
\frac{\lambda}{4!}\phi^4,
\label{e:eom_Veff}
\end{equation}
where the subscript ``$,\phi$'' denotes a partial derivative with respect to $\phi$.
At low densities $\rho/M^2 \ll \mu^2$, $\Veff$ has a local maximum at $\phi=0$ as well as two minima $\phi = \pm \phibr$ with 
\begin{equation}
\phibr = \mu\sqrt{6/\lambda}, 
\quad
\mbr \equiv \ddVeff^{1/2}(\phibr) = \mu\sqrt{2}. 
\label{e:broken_symmetry}
\end{equation} 
The $\mathbb{Z}_2$ symmetry $\phi \rightarrow -\phi$ is spontaneously broken as the field chooses one of these minima.  Henceforth we assume $\phi = \phibr$ in the broken-symmetry phase.  Meanwhile, at high densities, the mass-squared term  $m_0^2 = \rho/M^2 - \mu^2$ 
is positive, and $\phi=0$ is the only minimum of the potential. 

Consider an object with $\rho/M^2 \gg \mu^2$ at rest in a vacuum.  If the object is sufficiently small, we may linearize about the VEV, $\phi = \phibr + \delta\phi$, reducing (\ref{e:eom_Veff}) to 
$
\nabla^2 \delta\phi 
\approx
\rho \phibr / M^2 + 2 \mu^2 \delta\phi.
$
For small $\mu$ this looks like Poisson's equation $\nabla^2 \Psi = \rho / (2\Mpl^2)$ for the gravitational potential $\Psi$, so we have $\delta\phi \approx 2\Mpl^2M^{-2} \phibr \Psi$.  Evidently $\delta\phi$ couples linearly to the density with an effective matter coupling $\bbr \equiv \phibr\Mpl/M^2 = 6.0 \, \lambda^{-1/2} (\mu/ 10^{-3}\textrm{ eV}) (M/\textrm{TeV})^{-2}$ which is much smaller than the apparent coupling $\beta \equiv \Mpl/M = 2\times 10^{15} (M/\textrm{TeV})^{-1}$.  The symmetron in the linear regime behaves like a Yukawa scalar with mass $\mbr$ and force strength $\alpha = 2 \bbr^2$.  Even this attenuated coupling $\bbr$ is highly constrained by short-range gravity experiments, which exclude $\alpha \gtrsim 0.1$ for $\mbr^{-1} \gtrsim 0.1$~mm.  Note that increasing $\lambda$ decreases $\phibr$ and hence $\bbr$.  We will see that this qualitative behavior extends to torsion pendulum experiments, which place a lower bound on $\lambda$.

If we consider larger and larger objects, then the assumption of linearity is eventually violated; $\phi \approx 0$ deep inside the object, the ``source'' $\dVeff$ in (\ref{e:eom_Veff}) turns off, and the object becomes screened.  Since $\delta\phi$ cannot be less than $-\phibr$, linearity breaks down when $|\Psi| \gtrsim M^2/(2\Mpl^2)$.  A typical laboratory test mass with $\rho = 10$~\gcmc, size $\sim 1$~cm, and $|\Psi| \sim 3 \times 10^{-27}$  will be screened for $M \lesssim 100$~TeV.  It is this screened, nonlinear regime of the symmetron fifth force which we study here.  The fifth force on a test particle outside this screened object will be sourced only by a thin shell of matter near the surface of the object, inside which $0 \ll \phi \lesssim \phibr$.  Thus screening suppresses the fifth force.

At very low $M$, the symmetron-matter coupling $\beta = \Mpl/M$ is large, and symmetrons should be visible in colliders.  Reference~\cite{Brax_etal_2009} computed collider constraints on chameleon models and found that the coupling energy had to be greater than $\sim 1$~TeV.  Although a similar analysis for symmetrons is beyond the scope of this paper, it seems unlikely that $M \lesssim 100$~GeV is consistent with collider data.  Thus we are interested in the range $100\textrm{ GeV} \lesssim M \lesssim 100$~TeV.

We have also considered adding a photon coupling term $\phi^2 F_{\mu\nu}F^{\mu\nu} / (8\Mg^2)$ to (\ref{e:eom_Veff}) as in Ref.~\cite{Olive_Pospelov_2007}, making the symmetron accessible to oscillation experiments~\cite{Chou_2008,Chou_etal_2009,Steffen_etal_2010,Upadhye_Steffen_Chou_2012,Brax_Lindner_Zioutas_2012}.  An analysis similar to~\cite{Upadhye_Steffen_Weltman_2010} shows that symmetron-photon oscillation will occur in the broken-symmetry phase with an effective coupling $\bgbr = \phibr \Mpl / \Mg^2 = 6 \lambda^{-1/2} (\mu/10^{-3}\textrm{eV})(\Mg/\textrm{TeV})^{-2}$.  However, oscillation experiments  probe $\bgbr \gtrsim 10^{10}$ for $\mu \lesssim 10^{-2}$~eV, which requires either a small $\lambda$ strongly excluded by the fifth force constraints of Sec.~\ref{sec:constraints_on_symmetrons} or a small $\Mg$ likely excluded by colliders.

{{\bf{\em{B.~One-dimensional planar configurations: }}}
The symmetron field profile $\phi(z)$ can be found exactly for a planar gap between two thick planar slabs.  Let $\rho(z) = \rhom$ for $|z| \geq \Delta z / 2$, inside the slabs of matter, and $\rho(z) = \rhov$ for $|z| < \Delta z / 2$, the vacuum between the slabs.  Assume that $\rhov < \mu^2 M^2 < \rhom$, so that the matter is screened and the vacuum is possibly in the broken-symmetry phase. The equation of motion~(\ref{e:eom_Veff}) reduces to $\phi_{,zz} = \dVeff(\phi,\rho(z))$, where $\phi_{,zz} \equiv d^2\phi/dz^2$. Using $\frac{d^2}{dz^2} \phi = \frac{1}{2} \frac{d}{d\phi} (\frac{d\phi}{dz})^2$ to integrate the equation of motion over any interval $(z_A,z_B)$ with constant $\rho$, we have
\begin{equation}
\frac{\phi_{,z}(z_B)^2}{2} - \frac{\phi_{,z}(z_A)^2}{2}
=
\Veff(\phi_B,\rho) - \Veff(\phi_A,\rho),
\label{e:eom_integrated}
\end{equation}
with $\phi_A = \phi(z_A)$ and $\phi_B = \phi(z_B)$.

Equation~(\ref{e:eom_integrated}) is helpful if we choose either $z_A$ or $z_B$ such that $d\phi/dz = 0$.  Choosing the interval $(0,\Delta z / 2)$ we obtain ${\phis}_{,z}^2/2 = \Veff(\phis,\rhov) - \Veff(\phig,\rhov)$, where we have defined $\phig = \phi(0)$ to be the field at the center of the gap and $\phis = \phi(\Delta z/2)$ to be the field on the surface of one of the planar slabs. With the interval $(\Delta z / 2, \infty)$ we have $-{\phis}_{,z}^2/2 = -\Veff(\phis,\rhom),$ from which we obtain
\begin{equation}
\phis^2
=
\frac{\mu^2 M^2 - \rhov}{\rhom-\rhov}\phig^2 
- 
\frac{\lambda M^2}{12(\rhom-\rhov)} \phig^4.
\label{e:phis}
\end{equation}

Next we choose an interval $(0,z)$ for $0 < z < \Delta z / 2$ to determine $\phi_{,z}(z)/ \sqrt{2} = - \sqrt{\Veff(\phi,\rhov) - \Veff(\phig,\rhov)}$ inside the gap.  Integrating, we find $\phi(z)$ implicitly,
\begin{eqnarray}
\muv z
&=&
\int_\phi^{\phig} 
\frac{ \muv d\phi / \sqrt{2} }
     {\sqrt{\Veff(\phi,\rhov)-\Veff(\phig,\rhov)}}
\nonumber\\
&=&
\left(1-\frac{\phig^2}{2\phibr^2}\right)^{\!-\frac{1}{2}}
\!\!
\left[ 
  F\left(\frac{\pi}{2},\kg\!\right) 
  - 
  F\left(\!\sin^{-1}\frac{\phi}{\phig},\kg\!\right)
  \right]\,\quad
\label{e:Zgap}
\end{eqnarray}
where $\muv^2 \equiv \mu^2 - \rhov/M^2$, $\kg^2 \equiv \phig^2/(2\phibr^2-\phig^2)$, and $F(\theta,k) \equiv \int_0^\theta dt / \sqrt{1-k^2 \sin(t)^2}$ is the elliptic integral of the first kind.  We ``solve'' the gap by guessing $\phig$, using (\ref{e:phis}) to find $\phis$, using (\ref{e:Zgap}) with $\phi=\phis$ to find $z$, and refining our guess $\phig$ until $z = \Delta z / 2$.  Once the correct value of $\phig$ is known, (\ref{e:phis},\ref{e:Zgap}) determine $\phi$ everywhere in the gap.

The quadratic nature of the potential near $\phi=0$ implies that there is a minimum gap size $\Dzmin$ below which $\phi(z)=0$ everywhere.  We can see this by estimating the energy change due to pulling the field to some nonzero $\phig$ inside the gap.  The ``gradient energy'' density associated with this change in the field over a distance $\Delta z$ is of order $(\phig/\Delta z)^2$, while the potential energy density is of order $-\muv^2 \phig^2$.  Thus the net change in energy is negative only if $\Delta z \gtrsim \muv^{-1}$.  We can find the precise value of $\Dzmin$ by considering (\ref{e:Zgap}) with $\phi=\phis$ in the limit $\phig \rightarrow 0$,
\begin{equation}
\Dzmin
=
\frac{2}{\muv}
\left(\frac{\pi}{2} - \sin^{-1}\sqrt{\frac{\muv^2 M^2}{\rhom-\rhov}}\right).
\label{e:Dzmin}
\end{equation}
Typically $\rhom \gg \rhov$,~$\muv^2M^2$, so the second term inside the parenthesis can be neglected, and $\Dzmin \approx \pi/\muv$.

Finally, we determine $\phi(z)$ inside the slabs of matter.  Using (\ref{e:eom_integrated}) with interval $(z,\infty)$ and $z > \Delta z / 2$, we find $\phi_{,z} = -\sqrt{2\Veff(\phi,\rhom)} = -\sqrt{m_0^2 \phi^2 + \lambda\phi^4/12}$ with $m_0^2 = \rhom/M^2 - \mu^2$.  Defining $\varphis^2 = \lambda\phis^2/(12m_0^2)$ and integrating,
\begin{equation}
\frac{\phi(z)}{\phis}
=
\frac{2 e^{-m_0(z-\Delta z / 2)}(\sqrt{1+\varphis^2}-1)}
     {\varphis^2 - e^{-2m_0(z-\Delta z / 2)}(\sqrt{1+\varphis^2}-1)^2}.
\label{e:phi_slab}
\end{equation}


\section{Constraints on symmetrons}
\label{sec:constraints_on_symmetrons}

{{\bf{\em{A.~One-dimensional plane-parallel approximation: }}}
The previous section determined the surface field $\phis(\Delta z)$ for arbitrary gap size $\Delta z$, as well as the field profile $\phi(z)$ inside a planar slab of matter bounding the gap.  Rather than planar slabs, a torsion pendulum experiment such as \eotwash~uses parallel planar disks with surface features such as holes.  Let the $z$ axis be normal to both disks.  As a hole on the ``source'' disk moves past another hole on the ``test'' disk, fifth forces between the holes exert torques on the test disk.  
The 1Dpp approximation~\cite{Upadhye_2012a} estimates
the field at a point $(x,y)$ on the surface of each disk by $\phis(\Delta z(x,y))$, where $\Delta z(x,y)$ is the distance to the nearest point on the opposite disk.  The field inside the disk is approximated by (\ref{e:phi_slab}) given the surface field.  Since $\phis(x,y)$ will be greater for a region directly across from a hole on the opposite disk, there is an energy cost to moving the holes on opposite disks out of alignment with one another.  This change in energy as the source rotates is used to predict the torque signal. 

\begin{figure}[t]
\begin{center}
\includegraphics[angle=270,width=3.3in]{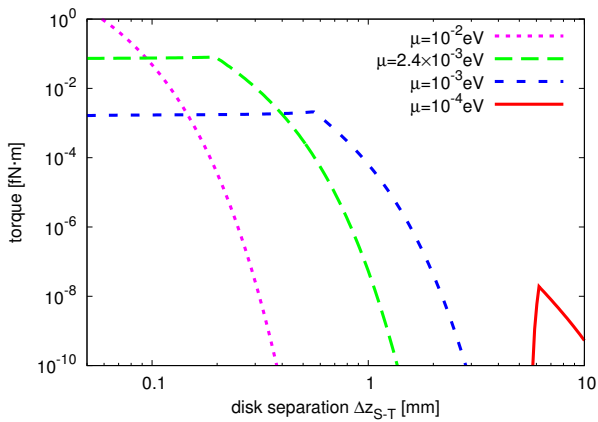}
\caption{Estimated torque in the apparatus of~\cite{Kapner_etal_2007} for several symmetron models.  $\lambda=1$ for all models shown; $M=500$~GeV for $\mu=10^{-2}$~eV and $M=1$~TeV otherwise. \label{f:torque_vs_DzST}}
\end{center}
\end{figure}

We find the energy per unit area associated with a point $(x,y)$ on the surface of a disk by integrating over the interior, with the $z$-dependence of $\phi$ given by (\ref{e:phi_slab}):
\begin{equation}
\frac{E}{A}
=
\int_\frac{\Delta z}{2}^\infty \left(\frac{\phi_{,z}^2}{2} + \Veff\right) dz
=
2 \int_\frac{\Delta z}{2}^\infty \Veff(\phi,\rhom) \, dz.
\label{e:energy_per_area}
\end{equation}
Consider a region on the test disk across from a hole of radius $\rS$ on the source disk.  The total energy of the field in this region is found by integrating (\ref{e:energy_per_area}), $\ET = \int_0^{\rS} 2\pi r \frac{E}{A} dr$.  We can similarly find the energy $\ES$ in the region of the source disk across from a hole of radius $\rT$ on the test disk.  Let $\DET$ be the energy difference between $\ET$ and the corresponding energy for a region on the test disk not overlapping a source disk hole, and define $\DES$ analogously for the source disk.  Then the total energy cost associated with moving the source and test holes out of alignment is $\DES + \DET$.  
If the disks have $\Nholes$ holes in $\Nrows$ rows, then the torque is
\begin{equation}
\tau
=
\frac{\Nholes^2}{2\Nrows} \fscreen (\DES + \DET)
\label{e:torque}
\end{equation}
where $\fscreen \sim \exp(-m_0 \zfoil)$ for $\zfoil = 10$~$\mu$m accounts for the  shielding foil between source and test disks~\cite{Upadhye_2012a}.
This 1Dpp prediction for the signal in a torsion pendulum experiment is shown in Figure~\ref{f:torque_vs_DzST} for several models.

{{\bf{\em{B.~Torsion pendulum constraints: }}}
\eotwash~\cite{Kapner_etal_2007} looked for fifth forces using two molybdenum disks, of density $\rhom = 10$~\gcmc, in a $10^{-6}$~torr vacuum corresponding to $\rhov \sim 10^{-12}$~\gcmc.  Each disk had $\Nholes=42$ holes in $\Nrows=2$ rows.  Radii of the source and test disk holes were $\rS=1.6$~mm and $\rT=2.4$~mm, respectively. 

Here we estimate constraints from an \eotwash-like experiment which excludes torques (\ref{e:torque}) greater than $0.01$~fN$\cdot$m.  \eotwash~probed fifth forces over a range of source-test distances $0.05\textrm{ mm} < \DzST < 10$~mm, with the strongest constraints typically coming from the shortest distances.  For $\mu \gtrsim 10^{-3}$~eV we use fifth force bounds at $\DzST = 0.1$~mm.  At smaller $\mu$, the minimum gap size $\Dzmin$ can be several millimeters, so no symmetron fifth forces are predicted at the smallest $\DzST$. Thus we use $\DzST=6.5$~mm for constraints on $\mu=10^{-4}$~eV. 

\begin{figure}[t]
\begin{center}
\includegraphics[angle=270,width=3.3in]{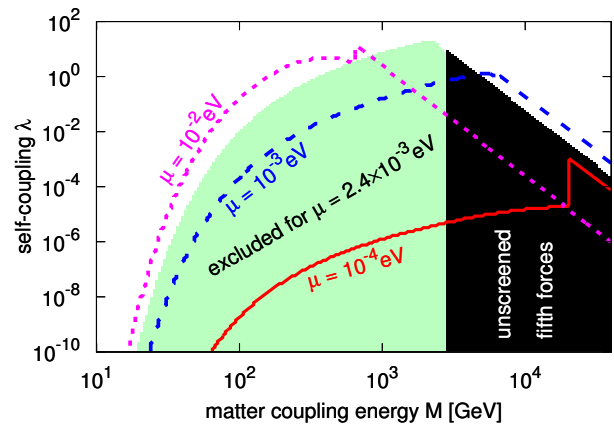}
\caption{Estimated constraints on symmetron dark energy.  The light-green shaded region is excluded for $\mu = M_\Lambda = 2.4 \times 10^{-3}$~eV, while models in the black shaded region are unscreened.  The solid red, dashed blue, and dotted purple curves show lower bounds on $\lambda$ for $\mu=10^{-4}$~eV, $10^{-3}$~eV, and $10^{-2}$~eV, respectively. \label{f:constraints}}
\end{center}
\end{figure}

For $\mu = M_\Lambda$ and $M=1$~TeV, we find that $\lambda < 7.5$ is excluded.
Figure~\ref{f:constraints} estimates constraints on  a wide range of symmetron models.  The 1Dpp calculation likely underestimates actual constraints by a factor of $2$-$3$, since only one $\DzST$ is used for each model, and since (\ref{e:energy_per_area}) only counts the energy of the field inside the disks~\cite{Upadhye_2012a}.  However, we do not attempt to correct for this here.

At $\mu  < 10^{-4}$~eV, $\Dzmin$ is too large to be probed by \eotwash, though constraints from larger-scale experiments apply.  For $\mu \gg 10^{-2}$~eV, matter becomes unscreened at all $M$ of interest.
At low $M$ the Compton wavelength is small, and screening by the shielding foil~\cite{Mota_Shaw_2006} weakens constraints.
Meanwhile, at large $M$, $\rhom < \mu^2 M^2$ and the gravitational potential of each disk $\Psi_\mathrm{disk} \ll M^2/(2\Mpl^2)$.  
Determining precise constraints in this linear regime is beyond the scope of this work, but such  models tend to have large fifth forces.


\section{Conclusion}
\label{sec:conclusion}

Symmetron dark energy is an intriguing new model in which fifth forces between massive objects are screened through a symmetry restoration at high densities.  We have considered a previously unexplored parameter region in which the symmetron self-coupling $\lambda \sim 1$, the vacuum mas $\mu \sim M_\Lambda$ so that only one small scale is necessary, and the matter coupling energy $M \sim 1$~TeV could be associated with new physics beyond the Standard Model.  Since symmetrons in this regime give rise to fifth forces at distances $\sim 0.1$~mm, they can be tested by current submillimeter fifth force experiments.

We have found the exact field profile (\ref{e:phis}-\ref{e:phi_slab}) in a planar geometry.  Using the 1Dpp approximation, we have predicted the torque signal due to the symmetron in an \eotwash-like torsion pendulum experiment.  Figure~\ref{f:torque_vs_DzST} shows this prediction for several models and experimental configurations, and the resulting constraints are shown in Fig.~\ref{f:constraints} for a large range in parameter space.  Specifically, we have shown that $\lambda<7.5$ is excluded for $\mu=M_\Lambda$ and $M=1$~TeV, demonstrating the power of existing experiments to probe symmetron dark energy.  These constraints may be improved in the near future by using a numerical computation such as that of~\cite{Upadhye_Gubser_Khoury_2006} and analyzing data from the next-generation \eotwash~Experiment.

{{\bf{\em{Acknowledgments:}}}
We are grateful to C.~Burrage, S.~Habib, K.~Heitmann, J.~Khoury, J.~Steffen, and especially to E.~Adelberger, for insightful discussions and helpful suggestions.
The author was supported by the U.S. Department of Energy, Basic Energy Sciences, Office of Science, under contract No. DE-AC02-06CH11357.

The submitted manuscript has been created by
UChicago Argonne, LLC, Operator of Argonne
National Laboratory (``Argonne''). Argonne, a
U.S. Department of Energy Office of Science laboratory,
is operated under Contract No. DE-AC02-
06CH11357. The U.S. Government retains for itself,
and others acting on its behalf, a paid-up
nonexclusive, irrevocable worldwide license in said
article to reproduce, prepare derivative works, distribute
copies to the public, and perform publicly
and display publicly, by or on behalf of the Government.

\bibliographystyle{unsrt}
\bibliography{chameleon}

\begin{thebibliography}{10}

\bibitem{Komatsu_etal_2010}
E.~Komatsu et~al.
\newblock {\em Astrophys.J.Suppl.}, 192:18, 2011.

\bibitem{Larson_etal_2011}
D.~Larson et~al.
\newblock {\em Astrophys.~J.~Suppl.}, 192:16, 2011.

\bibitem{Suzuki_etal_2012}
N.~Suzuki et~al.
\newblock {\em Astrophys.~J.}, 746:85, 2012.

\bibitem{Sanchez_etal_2012}
Ariel~G. Sanchez et~al.
\newblock {\em Mon.~Not.~R.~Astron.~Soc.}, 425:415, 2012.
\newblock e-Print arXiv:1203.6616.

\bibitem{Khoury_Weltman_2004a}
J.~Khoury and A.~Weltman.
\newblock {\em Phys.~Rev.~Lett.}, 93:171104, 2004.

\bibitem{Khoury_Weltman_2004b}
J.~Khoury and A.~Weltman.
\newblock {\em Phys. Rev. D}, 69:044026, 2004.

\bibitem{Brax_etal_2004}
Ph. Brax, C.~van~de Bruck, A.-C. Davis, J.~Khoury, and A.~Weltman.
\newblock {\em Phys. Rev. D}, 70:123518, 2004.

\bibitem{Vainshtein_1972}
A.~I. Vainshtein.
\newblock {\em Phys.~Lett.~B}, 39:393, 1972.

\bibitem{Dvali_Gabadadze_Porrati_2000}
G.~R. Dvali, G.~Gabadadze, and M.~Porrati.
\newblock {\em Phys. Lett. B}, 485:208--214, 2000.

\bibitem{Deffayet_etal_2001}
C.~Deffayet et~al.
\newblock {\em Phys.Rev.}, D65:044026, 2002.

\bibitem{Nicolis_Rattazzi_Trincherini_2008}
A.~Nicolis, R.~Rattazzi, and E.~Trincherini.
\newblock {\em Phys.Rev.}, D79:064036, 2009.

\bibitem{Hinterbichler_Khoury_2010}
K.~Hinterbichler and J.~Khoury.
\newblock {\em Phys.~Rev.~Lett.}, 104:231301, 2010.

\bibitem{Olive_Pospelov_2007}
K.~A. Olive and M.~Pospelov.
\newblock {\em Phys.Rev.}, D77:043524, 2008.

\bibitem{Adelberger_Heckel_Nelson_2003}
E.~G. Adelberger, B.~R. Heckel, and A.~E. Nelson.
\newblock {\em Ann.~Rev.~Nucl.~Part.~Sci.}, 53:77--121, 2003.

\bibitem{Adelberger_etal_2009}
E.~G. Adelberger, J.~H.~Gundlach nd~B.~R.~Heckel, S.~Hoedl, and
  S.~Schlamminger.
\newblock {\em Prog. Part. Nucl. Phys.}, 62:102--134, 2009.

\bibitem{Upadhye_Gubser_Khoury_2006}
A.~Upadhye, S.~S. Gubser, and J.~Khoury.
\newblock {\em Phys.~Rev.~D}, 74:104024, 2006.

\bibitem{Brax_etal_2007c}
P.~Brax, C.~van~de Bruck, A.~C. Davis, D.~F. Mota, and D.~J. Shaw.
\newblock {\em Phys.~Rev.~D}, 76:124034, 2007.
\newblock e-Print arXiv:0709.2075.

\bibitem{Brax_etal_2010b}
P.~Brax, C.~van~de Bruck, A.-C. Davis, D.~J. Shaw, and D.~Iannuzzi.
\newblock {\em Phys.~Rev.~Lett.}, 104:241101, 2010.

\bibitem{Kapner_etal_2007}
D.~J. Kapner, T.~S. Cook, E.~G. Adelberger, J.~H. Gundlach, B.~R. Heckel, C.~D.
  Hoyle, and H.~E. Swanson.
\newblock {\em Phys.~Rev.~Lett.}, 98:021101, 2007.
\newblock e-Print arXiv:hep-ph/0611184.

\bibitem{Upadhye_2012a}
A.~Upadhye.
\newblock {\em Phys.~Rev.~D}, 86:102003, 2012.
\newblock e-Print: arXiv:1209.0211.

\bibitem{Hinterbichler_Khoury_Levy_Matas_2011}
K.~Hinterbichler, J.~Khoury, A.~Levy, and A.~Matas.
\newblock {\em Phys. Rev. D}, 84:103521, 2011.

\bibitem{Brax_etal_2009}
P.~Brax, C.~Burrage, A.-C. Davis, D.~Seery, and A.~Weltman.
\newblock {\em JHEP}, 0909:128, 2009.
\newblock e-print arXiv:0904.3002.

\bibitem{Chou_2008}
A.~S. Chou et~al.
\newblock {\em Phys.~Rev.~Lett.}, 100:080402, 2008.
\newblock ePrint: arXiv:0710.3783.

\bibitem{Chou_etal_2009}
A.~S. Chou, W.~C. Wester, A.~Baumbaugh, H.~R. Gustafson, Y.~Irizarry-Valle,
  P.~O. Mazur, J.~H. Steffen, R.~Tomlin, A.~Upadhye, A.~Weltman, X.~Yang, and
  J.~Yoo.
\newblock {\em Phys. Rev. Lett}, 102:030402, 2009.

\bibitem{Steffen_etal_2010}
J.~H. Steffen et~al.
\newblock {\em Phys.~Rev.~Lett.}, 105:261803, 2010.
\newblock ePrint: arXiv:1010.0988.

\bibitem{Upadhye_Steffen_Chou_2012}
A.~Upadhye, J.~H. Steffen, and A.~S. Chou.
\newblock {\em Phys. Rev. D}, 86:035006, 2012.

\bibitem{Brax_Lindner_Zioutas_2012}
P.~Brax, A.~Lindner, and K.~Zioutas.
\newblock {\em Phys.~Rev.~D}, 85:043014, 2012.

\bibitem{Upadhye_Steffen_Weltman_2010}
A.~Upadhye, J.~H. Steffen, and A.~Weltman.
\newblock {\em Phys.~Rev.~D}, 81:015013, 2010.

\bibitem{Mota_Shaw_2006}
D.~F. Mota and D.~J. Shaw.
\newblock {\em Phys. Rev. Lett.}, 97:151102, 2006.

\end{thebibliography}
\end{document}